\documentclass[twocolumn,showpacs,preprintnumbers,amsmath,amssymb,superscriptaddress]
{revtex4}
\usepackage[english]{babel}
\usepackage{amsmath,amssymb}
\usepackage{graphicx}

\large
\begin{document}

\title{Electrokinetic Instability near Charge-Selective Hydrophobic Surfaces}
\author{V.~S. Shelistov}\email{shelistov_v@mail.ru}
\affiliation{Scientific Research Department, Kuban State Universaty, Krasnodar, 350040, Russian
Federation.}
\author{ E.~A. Demekhin} \email{edemekhi@gmail.com}
 \affiliation{Department of Computation Mathematics and Computer Science,
Kuban State University,
Krasnodar, 350040, Russian Federation.}
\affiliation{Laboratory of General Aeromechanics, Institute of Mechanics, 
Moscow State University, Moscow, 117192, Russian Federation.}
\author{G.~S. Ganchenko}\email{ganchenko.ru@gmail.com}
\affiliation{Department of Computation Mathematics and Computer Science,
Kuban State University,
Krasnodar, 350040, Russian Federation.}

\date{\today}

\begin{abstract}
The influence of the texture of a hydrophobic surface on the electro-osmotic slip of the
second kind and the electrokinetic instability near charge-selective surfaces
(permselective membranes, electrodes, or systems of micro- and nanochannels) is
investigated theoretically using a simple model based on the Rubinstein--Zaltzman
approach. A simple formula is derived to evaluate the decrease in the instability
threshold due to hydrophobicity. The study is complemented by numerical
investigations both of linear and nonlinear instabilities near a hydrophobic
membrane surface. Theory predicts a significant enhancement of the ion flux to the
surface and shows a good qualitative agreement with the available experimental data.
\end{abstract}

\maketitle
\section{Introduction}

%\textit{Motivation.}
The investigation of the wettability of solid materials has a history of at least two centuries; this interest has been caused by its fundamentals and its practical implications.
During the last decade, the interest in hydrophobic surfaces has increased
rapidly because of their applications to nanotechnology \cite{Vin1}. In
particular, such surfaces can dramatically change the transport properties of
the flow, and this is especially important in the micro- and nanoscales. The presence
of a special surface microtexture can drastically increase its hydrophobic
properties. Such a texture may be either natural or artificial.

One of the remaining challenges in this field is related to the possibility of
greatly enhancing transport properties by combining hydrophobicity with
electrokinetic phenomena. In Ref. \cite{Mul1} it is shown that the classical
Helmholtz--Smoluchowski electro-osmotic velocity acquires an additional factor of
$1+b/\lambda_D$, where $b$ is the slip length and $\lambda_D$ is the Debye
length.

Much more interesting is the case of an extended polarization near a
charge-selective surface (membrane, electrode or system of micro- and
nanochannels). Such a polarization can drive a hydrodynamic flow if a tangential
component of the electric field exists along with its normal component. Two key
mechanisms can be responsible for the formation of such a tangential field: a curvature  of the surface
(the so-called Dukhin mechanism), or an
electrokinetic instability (the mechanism of Rubinstein--Zaltzman)
(see \cite{Chang1,Chang2}). 

The effect of a coupling near a charge-selective
surface between electrokinetic phenomena and the hydrophobicity of such a surface is expected to be especially
strong. Such surfaces are known to be enormously sensitive to any microscopic
geometric alterations \cite{Chang2}, and is reasonable to expect a
strong sensitivity with respect to their hydrophobicity. Indeed, in the recent
experiments by Nikonenko et al. \cite{Nik1}, several cation-exchange membranes
with different degrees of hydrophobicity were investigated for the parameters
corresponding to electrokinetic instability. It was found that an increase
in the surface hydrophobicity is accompanied with a significant rise in
the overlimiting mass transfer and a decrease in the threshold of the instability.
A theoretical investigation of this practically important fundamental problem has been
lacking.

\textit{Two kinds of slip.}
It is known that on a hydrophobic surface that, instead of what is usual for Newtonian
liquids (where all velocity components vanish, both tangential and normal, i.e.,
$$
\mathbf{u}_s=0, \quad u_n=0),
$$
only the normal component vanishes while the tangential velocity components have a
slip along the surface, i.e.,
$$
\mathbf{u}_s=b \frac{\partial \mathbf{u}_s}{\partial n}, \quad u_n=0,
$$
which is characterized by the slip length $b$. The experimental information
about the value of $b$ is rather contradictory
%\cite{Mul1,Joly,Stone,Vin2,Cottin}
and it can be roughly evaluated as $b\leqslant100$~nm, depending on the wall
material. For natural superhydrophobicity, the slip length is much larger and
can be up to several microns. In both cases, the hydrophobicity and
the natural superhydrophobicity $b$ is a scalar. An artificially fabricated micro-
or nanotexture is usually oriented in some direction and, as a consequence,
for such anisotropic surfaces, $b$ is a $2\times2$ matrix \cite{Vin2,Vin3}.
However, by a suitable averaging, an anisotropic surface can be considered as an
isotropic one with some effective scalar slip $\langle b\rangle$.

A non-uniform tangential electric field forms an ion flux that is non-uniform along the membrane
surface. This, in turn, gives rise to another kind of slip. At a
distance of the space-charge region length, $50 \div 1000$~nm, the tangential
velocity rapidly changes from zero to
$$
\mathbf{u}_s \sim \Delta V^2 \frac{1}{j_n} \nabla_s{j_n},
$$
where $\Delta V$ is the potential drop in the space charge region, $j_n$ is the normal
to the surface ion flux, and $\nabla_s$ is the surface Hamilton operator
\cite{Rub15a}. In order to distinguish two slip velocities, we shall call the
latter one the `electro-osmotic slip velocity' and the first one, `hydrophobic slip' or
just `slip velosity'.

The electro-osmotic slip is a driving force to destabilize the flow in a diffusion
region if it prevails over the viscous dissipation. On the other hand, the
hydrophobicity and corrensponding hydrophobic slip results in a diminishing of
this dissipation.

\textit{Outline.}
In this paper, we present an analytical and numerical investigation of the coupling
between these two kinds of slip. The formulation of the problem is given in
Section II and is based on a binary dilute electrolyte model and creeping flow
assumption. Section III is focused on an investigation of linear stability in the
limit of a small Debye number. A~simple formula is derived to evaluate the decrease
in the instability threshold due to hydrophobicity. This investigation is complemented
in Section IV by numerical investigations both of the linear and nonlinear instabilities
near a hydrophobic membrane surface. Theory predicts a significant enhancement of
the ion flux to the surface and shows a good qualitative agreement with
the available experimental data.

\section{Statement}

A binary electrolyte between semi-selective ion-exchange membrane surfaces,
$\tilde{y}=0$ and $\tilde{y}=\tilde{L}$, is considered. The lower membrane
surface is assumed to be hydrophobic with a slip boundary condition while the
upper membrane obeys a regular no-slip condition. Tilded notations are used for
the dimensional variables, as opposed to their untilded dimensionless
counterparts. The diffusivities of the cations and anions are assumed to be equal,
$\tilde{D}^+=\tilde{D}^-=\tilde{D}$. The characteristic quantities to make the
system dimensionless are 
$\tilde{L}$, the characteristic length, the distance between the membranes; 
$\tilde{L}^2/\tilde{D}$, the characteristic time;
$\tilde{\mu}$, the dynamic viscosity, which is taken as a characteristic dynamical
value;
$\tilde{\Phi}_0 = \tilde{R}\tilde{T}/\tilde{F}$, the potential, which is taken
as characteristic;
$\tilde{c}_0$, the unperturbed bulk ion concentration of the one-dimensional (1D)
solution.
Then, the electroconvection is described by the equations for ion transport,
Poisson's equation for the electric potential, and Stokes's equation for a
creeping flow:
\begin{widetext}
\begin{equation}\label{eq1}
 \displaystyle\frac{\partial c^+}{\partial t}+ u\frac{\partial c^+}{\partial x}+
 v\frac{\partial c^+}{\partial y}
 = \phantom{-}\frac{\partial }{\partial x}\left(c^+
 \frac{\partial \Phi}{\partial x} \right) +
 \frac{\partial }{\partial y}\left(c^+ \frac{\partial \Phi}{\partial y}\right)+
 \frac{\partial^2 c^+}{\partial x^2}+
 \frac{\partial^2 c^+}{\partial y^2},
\end{equation}

\begin{equation}\label{eq2}
 \displaystyle\frac{\partial c^-}{\partial t}+ u\frac{\partial c^-}{\partial x}+v
 \frac{\partial c^-}{\partial y}= -\frac{\partial }{\partial x}\left(c^- \frac{\partial \Phi}{\partial y}\right)
 - \frac{\partial }{\partial y}\left(c^- \frac{\partial \Phi}{\partial y}\right)+
 \frac{\partial^2 c^-}{\partial x^2}+
 \frac{\partial^2 c^-}{\partial y^2},
\end{equation}
\end{widetext}
\begin{equation}\label{eq3}
 \nu^2 \left \{ \frac{\partial^2 \Phi}{\partial x^2}+
 \frac{\partial^2 \Phi}{\partial y^2}\right\}=c^--c^+,
\end{equation}

\begin{equation}\label{eq3a}
 -\frac{\partial{P} }{\partial {x}} +
 \frac{\partial^2 {u}}{\partial {x}^2}+
 \frac{\partial^2 {u}}{\partial {y}^2}=\frac{\varkappa}{\nu^2}
\left({c} ^+-{c} ^-\right)\frac{\partial {\Phi}}{\partial {x}},
\end{equation}
\begin{equation}\label{eq3b}
 -\frac{\partial{P} }{\partial {y}} +
 \frac{\partial^2 {v}}{\partial{x}^2}+
 \frac{\partial^2{v}}{\partial{y}^2}=\frac{\varkappa}{\nu^2}
 \left({c} ^+-{c} ^-\right)\frac{\partial {\Phi}}{\partial {y}},
\end{equation}
\begin{equation}\label{eq3c}
 \frac{\partial {u}}{\partial {x}}+
 \frac{\partial {v}}{\partial {y}}=0.
\end{equation}

This system of dimensional equations is complemented by the proper boundary
conditions on the surface of ideal permselective membranes, which are as follows:
\begin{eqnarray}
 y=0: \quad &c^+=p, \quad &-c^-\frac{\partial \Phi}{\partial y}+
 \frac{\partial c^-}{\partial y}=0,
  \nonumber\\ \Phi=0,\quad
&v=0,\quad &u=b \frac{\partial u}{\partial y} 
 \label{eq7}
\end{eqnarray}
\begin{eqnarray}\label{eq8}
 y=1: \quad &c^+=p, \quad &-c^-\frac{\partial \Phi}{\partial y}+
 \frac{\partial c^-}{\partial y}=0,\nonumber\\ \Phi=\Delta V,\quad
 &v=0,\quad &u=0.
 \quad 
\end{eqnarray}

Conditions \eqref{eq7}--\eqref{eq8} are complemented by the relation
\begin{equation}\label{eq8a}
\int_0^1 c^-dy=1,
\end{equation}
which specifies the amount of negative ions in the 1D solution.
Here, ${c}^+$ and ${c}^-$ are the concentrations of the cations and anions,
respectively; a two-dimensional case is considered, $\mathbf{u}=\{{u},\:{v}\}$ is
the fluid velocity vector; $\{{x},\:{y}\}$ are the coordinates, $x$ is directed\
along the membrane surface, $y$ is normal to the membrane surface; $\Phi$ is the
electrical potential; $\Delta V$ is the potential drop between the membranes;
$P$ is the pressure; the interface concentration $c^+$ is equal to that of the
the fixed charges inside the membrane and the relation is asymptotically valid at $p \gg 1$ (see
\cite{Rub13}); $b$ is the slip length; $\nu$ is the dimensionless Debye
length, which is a small parameter,
$$
\nu=\frac{\tilde{\lambda}_D}{\tilde{L}}, \qquad \tilde{\lambda}_D=\left(\frac{\tilde{\varepsilon}\tilde{\Phi}_0}{\tilde{F}\tilde{c}_0}\right)^{1/2}=
\left(\frac{\tilde{\varepsilon}\tilde{R}\tilde{T}}{\tilde{F}^2\tilde{c}_0}\right)^{1/2},
$$
$\varkappa=\tilde{\varepsilon}\tilde{\Phi}_0^2/\tilde{\mu}\tilde{D}$ is a
coupling coefficient between the hydrodynamics and the electrostatics. It characterizes
the physical properties of the electrolyte solution and is fixed for a given liquid and
electrolyte. $\tilde{F}$ is Faraday's constant; $\tilde{R}$ is the universal gas
constant; $\tilde{T}$ is the temperature in Kelvin; and $\tilde{\varepsilon}$ is the
permittivity of the medium.

The parameters of the system are $\nu$, $\Delta V$, $\varkappa$ and $b$.
Because of the well-known weak dependence of the solution on $p$, it is fixed at
$p=5$ and $p$ is not included in the list of parameters.

The spactial average of the electric current through the semi-selective surface $y=0$ is
determined by only positive ions,
\begin{equation}\label{eq9}
 j =\frac{1}{l} \int_0^l\left( c^+ \frac{\partial \Phi}{\partial y} +\frac{\partial c^+}{\partial y}\right) dx,
\end{equation}
where $l$ is the length of the surface.

For some of our calculations, we shall use the following equation instead of
\eqref{eq3a}--\eqref{eq3c}:
\begin{eqnarray}\label{eq10}
 \frac{\partial^4 \Psi}{\partial x^4}&+&2
 \frac{\partial^4 \Psi}{\partial x^2 \partial y^2}+
 \frac{\partial^4 \Psi}{\partial y^4}=\nonumber\\
 &&\frac{\varkappa}{\nu^2} \left \{\frac{\partial}{\partial y}\left(\rho \frac{\partial \Phi}{\partial x}\right)-
 \frac{\partial}{\partial x}\left(\rho \frac{\partial \Phi}{\partial y}\right)\right\}
\end{eqnarray}
for the stream function $\Psi$, $u=\partial \Psi/\partial y$,
$v=-\partial\Psi/\partial x$.

\section{Asymptotic solution}

\textit{Correction to the electro-osmotic slip.}
For the 1D steady-state solution, $\partial/\partial t = \partial/\partial x = 0$,
Equations \eqref{eq1}--\eqref{eq8} turn into one nonlinear ODE (see
\cite{Bab1} and \cite{Rub16}):
\begin{equation}\label{eq19}
\quad \nu^2 \frac{d^2 E}{dy^2} + \Big[j(y_m-y)- \frac{\nu^2}{2}E^2 \Big]E+j=0.
\end{equation}
Here, $E \equiv \partial \Phi/\partial y$ and $y_m$ is the length of the
space-charge region (SCR). Hydrodynamic motion is not involved in this solution
because there is no tangential component of the electric field,
$\partial \Phi/\partial x = 0$. As $\nu \to 0$ and $E=O(1/\nu)$, $d/dy=O(1)$,
the solution of Eq. \eqref{eq19} in the SCR, $0 < y < y_m$, is
\begin{eqnarray}\label{eq21}
\quad \nu E= \nu \frac{\partial \Phi}{\partial y}=\sqrt{2j(y_m-y)},\nonumber\\
\nu\Phi=\frac{2\sqrt{2}}{3j}(jy_m)^{3/2}-\frac{(2jy_m-2jy)^{3/2}}{3j}.
\end{eqnarray}
From this relation we can get the length $y_m$ of the SCR,
\begin{equation}\label{eq23}
 y_m=\frac{9^{1/3}\nu^{2/3}{\Delta V_1}^{2/3}}{2j^{1/3}},
\end{equation}
where $\Delta V_1$ is the potential drop in the SCR.

Let us assume that the independent variables of the system in the SCR are not
constant with respect to $x$ but instead are slowly varying functions,
$\partial/\partial x \ll \partial/\partial y$, $v \ll u$. Then Equations
\eqref{eq3a}--\eqref{eq3c} turn into the following (see \cite{Rub13,Rub16}):
\begin{equation}\label{eq27}
\frac{\partial P}{\partial x}=\frac{\partial^2 U}{\partial
y^2}+\varkappa\frac{\partial E}{\partial y}\frac{\partial
\Phi}{\partial x}, \quad \frac{\partial P}{\partial y}=\varkappa E \frac{\partial E}{\partial y},
\end{equation}
with the boundary conditions
\begin{equation}\label{eq28a}
y = 0: \quad U = b\frac{dU}{dy}, \qquad
y = y_m: \quad \frac{\partial U}{\partial y} = 0.
\end{equation}
Upon excluding pressure from this system, integrating twice, and using the boundary
conditions \eqref{eq28a}, we obtain the electro-osmotic slip,
\begin{eqnarray}\label{eq31}
\frac{1}{\varkappa}U_m =
 -\frac{1}{8}(1&+&3\beta)\Delta V_1^2\frac{1}{j}\frac{\partial j}{\partial x}-\nonumber\\
&&(1+\frac{3}{2}\beta)\Delta V_1\frac{\partial \Delta V_1}{\partial x}, \\
\beta &\equiv& \frac{b}{y_m}.\nonumber
\end{eqnarray}
At $\beta=0$ this turns into the famous Rubinstein--Zaltzman formula
\cite{Rub15a}, which is valid for $\Delta V > O(|\log {\nu}|)$.

Neglecting the potential drop in the electroneutral diffusion part,
$\Delta V=\Delta V_1$, and assuming that all the potential drop across the
membrane is located in the space-charge region, we can find a simplified version
of the electro-osmotic slip:
\begin{equation}\label{eq31a}
U_m = -\frac{\varkappa}{8}(1+3\beta)\Delta V^2\frac{1}{j}
\frac{\partial j}{\partial x}.
\end{equation}

\textit{The diffusion region and critical points of instability.}
In the electroneutral diffusion region, $1 < y < y_m$, Equations
\eqref{eq1}--\eqref{eq3} and \eqref{eq8a} and \eqref{eq10} along with the boundary
conditions taken from the solution in the SCR turn into
\begin{equation}\label{eq35}
 \frac{\partial^4 \Psi}{\partial x^4}+2\frac{\partial^4
\Psi}{\partial x^2 \partial^2 y}+\frac{\partial^4 \Psi}{\partial
y^4}=0,
 \end{equation}
\begin{eqnarray}\label{eq36}
 y=y_m: \quad&\Psi=0, \quad&\frac{\partial \Psi}{\partial y}=U_m, \nonumber\\
 y=1:\quad&\Psi=0, \quad&\frac{\partial \Psi}{\partial y}=0.
 \end{eqnarray}
\begin{equation}\label{eq39}
 \frac{\partial K}{\partial t}+U\frac{\partial K}{\partial x}+V\frac{\partial K}{\partial
 y}=\frac{\partial ^2 K}{\partial x^2}+\frac{\partial ^2 K}{\partial y^2},
\end{equation}

\begin{eqnarray}\label{eq40}
 y&=&y_m:\quad K=a\left(\nu \frac{\partial K}{\partial y}\right)^{2/3},\nonumber\\
  y&=&1:\quad\int_{y_m}^1 K dy=2.
\end{eqnarray}

\begin{equation}\label{eq41a}
 \frac{\partial }{\partial x}\left(K\frac{\partial \Phi}{\partial x}\right)+ \frac{\partial }{\partial y}\left(K\frac{\partial \Phi}{\partial y}\right)=0,
\end{equation}
\begin{equation}\label{eq41b}
 y=y_m:\quad \Phi=\Delta V_1, \quad y=1:\quad \Phi=\Delta V.
\end{equation}
Here, $K=c^++c^-$ is the concentration of salt and $a \approx 0.966$. We have chosen the formulation by by Rubinstein and Zaltzman \cite{Rub16,Rub15a}, with some small
alterations. 

For the 1D steady state, $\partial/\partial t = \partial/\partial x = 0$, the
solution of \eqref{eq35}--\eqref{eq41b} is
\begin{equation}\label{eq41d}
y_m=
\Big(1-\frac{2}{\sqrt{j}}\Big)+\frac{a\nu^{2/3}}{j^{1/3}},
\end{equation}
\begin{equation}\label{eq41e}
K=\nu^{2/3}j^{2/3}+\frac{y-y_m}{1-y_m}\Big(-2a\nu^{2/3}j^{2/3}+\frac{4}{1-y_m}\Big),
\end{equation}
\begin{equation}\label{eq41f}
\Phi=\ln[1+\frac{j^{1/3}}{a\nu^{2/3}}(y-y_m)]+\frac{(2jy_m)^{2/3}}{3j\nu}.
\end{equation}
The terms of $O(\nu^{4/3})$ are neglected in the solution. Substitution of
$\Phi=\Delta V$ at $y=1$ into \eqref{eq41f} gives the volt--current
characteristic.

To study the linear stability of the solution found, we superimpose on it small
sinusoidal perturbations of the form $f=f_0(y)+\hat{f}(y)\exp{(ikx+\lambda t)}$,
$\hat{f} \to 0$. Linearizing \eqref{eq35}--\eqref{eq41b} with respect to
the perturbations turns the system into a system of linear ODEs, which can be
readily solved analytically. Only the case of marginal stability is considered,
$\lambda=0$; the neutral parameters are connected by the relation

\begin{eqnarray} \label{eq58a}
\frac{\varkappa}{8}&\cdot&\Delta V^2 \cdot(1+3\,\beta)=\nonumber\\
&&4\,\frac{(k^2-\sinh^2{k})\,(\sinh{k}+N\cdot k\cosh{k})}{k^3\,\cosh{k}-\sinh^3{k}}
\end{eqnarray}

with
$$
N=\frac{\nu^{2/3}}{3}\left\{\frac{2a}{j^{1/3}}+
\frac{2^{1/3}}{j^{1/3}}\cdot\left(\frac{3}{4}\,\Delta
 V\right)^{2/3}\right\}\;\mbox{and} \;h=1-y_m.
$$

The critical parameters of the threshold of instability are determined by the
condition
\begin{equation}\label{eq59}
\frac{\partial \Delta V}{\partial k}|_{k=k_*}=\infty,
\end{equation}
which gives
\begin{widetext}
\begin{equation}\label{eq60}
\frac{\varkappa}{32}\cdot(\Delta V^*)^2\left(1+3\beta \right) = \frac{\sinh{k^*}\cosh{k^*}-k^*+N\cdot k^*\,\left(\cosh^2{k^*}-k^*\,\coth{k^*}\right)}
{\sinh{k^*}\cosh{k^*}+k^*-2(k^*)^2\coth{k^*}}.
\end{equation}
\end{widetext}

The fraction in the right-hand side of \eqref{eq60} for small $\nu$ 
depends weakly on $k^*$ for its realistic values: indeed, changing the critical wave
number in the window $k^*=3 \div 8$ results in the fraction's changing from 1.06
to 1.0001. By using this fact, we re-write \eqref{eq60} as a very simple
relation,
\begin{equation}\label{eq61}
\frac{\Delta V^*}{\Delta V^*_0}=\frac{1}{\sqrt{1+3\beta}},
\end{equation}
where $\Delta V^*_0$ is taken at $\beta=0$.

\section{Numerical solution and discussion}
These analytical results will now be verified and complemented by numerical calculations of
the linear stability and direct numerical simulation of the system
\eqref{eq1}--\eqref{eq8a}.

\textit{Linear stability.}
Let us consider sinusoidal perturbations with a wave number $k$ superimposed on
the 1D steady-state solution,
$$
 c^{\pm}=c_0^{\pm}+\hat{c}^{\pm} \exp{(i k x+\lambda t)},
 $$
 $$
 \Phi=\Phi_0+\hat{\Phi} \exp{(i k x-\lambda t)}, \quad \Psi=\hat{\Psi} \exp{(i k x-\lambda t)}.
$$
The subscript $0$ is related to the mean solution; hat, to the perturbations. Upon
linearizing \eqref{eq1}--\eqref{eq8} with respect to the perturbations and
skipping the subscript $0$ in the mean solution, we get
\begin{eqnarray}\label{eq66}
\lambda \hat{c}^{+}-i k {D\hat{c}}^{+}\hat{\Psi} =\frac{d}{d y}\left( c^{+}D\hat{\Phi}+E\hat{c}^{+}+D\hat{c}^{+} \right)\nonumber\\
-k^2 c^+ \hat{\Phi}-k^2\hat{c}^{+},
\end{eqnarray}
\begin{eqnarray}\label{eq67}
\lambda \hat{c}^{-}-i k {D\hat{c}}^{-}\hat{\Psi} =\frac{d}{d y}\left(- c^-D\hat{\Phi}-E\hat{c}^{-}+D\hat{c}^{-} \right)\nonumber\\-k^2 c^- \hat{\Phi}-k^2\hat{c}^{-},
\end{eqnarray}
\begin{equation}\label{eq68}
\nu^2 \left( D^2\hat{\Phi} -k^2 \hat{\Phi }\right) = -\hat{\rho},
\end{equation}
\begin{equation}\label{eq69}
D^4\hat{\Psi}-2 k^2 D^2 \hat{\Psi} +k^4 \hat{\Psi}= i k
\frac{\kappa}{\nu^2}\left(D \rho\hat{\Phi}-E\hat{\rho} \right),
\end{equation}
\begin{eqnarray}\label{eq70}
y=0: &\;& \hat{\Phi}=0, \; \hat{\Psi}=0,\; \:D\hat{\Psi}=\beta y_mD^2\hat{\Psi}, 
 \nonumber\\ &&c^-D\hat{\Phi}+E\hat{c}^{-}-D\hat{c}^{-}=0, \; \hat{c}^{+}=0
\end{eqnarray}
\begin{eqnarray}\label{eq71}
y=1: &\;& \hat{\Phi}=0,\; \hat{\Psi}=\hat{\Psi^{'}}=0,\nonumber\\
&&c^-D\hat{\Phi}+E\hat{c}^{-}-D\hat{c}^{-}=0,\;
\hat{c}^+=0.
\end{eqnarray}
Here, $D$ means the derivative with respect to $y$, and $\rho=c^+-c^-$.

The Galerkin pseudo-spectral method with Chebyshev polynomials taken as the basic
functions \cite{O1} was employed to discretize the nonlinear equation \eqref{eq19}
and the eigenvalue problem \eqref{eq66}--\eqref{eq71}. The system of nonlinear
algebraic equations originating from \eqref{eq19} was then solved by the Newtonian method,
while the generalized matrix eigenvalue problem was solved by the QR-algorithm.
The number of functions was up to 512.

\begin{figure}[hbt]
\centering
\includegraphics[width=0.5\textwidth]{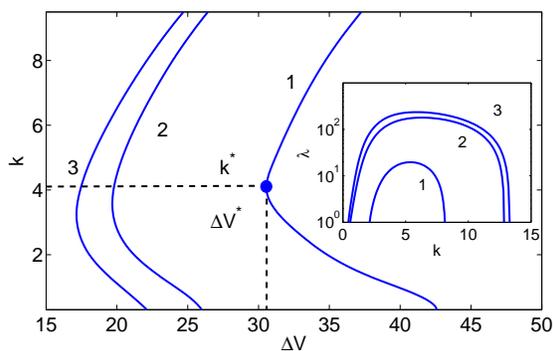}
\caption{(Color online) Marginal stability curves for $\varkappa=0.1$ and $\nu=10^{-3}$. 1 --- $\beta=0$; 2 --- $\beta=1$; 3 --- $\beta=3$. Inset: Increments of linear growth rate $\lambda$ versus wave number $k$ for different hydrophobicities $\beta$. }
\label{margi}
\end{figure}

\begin{figure}[hbt]
\centering
\includegraphics[width=0.5\textwidth]{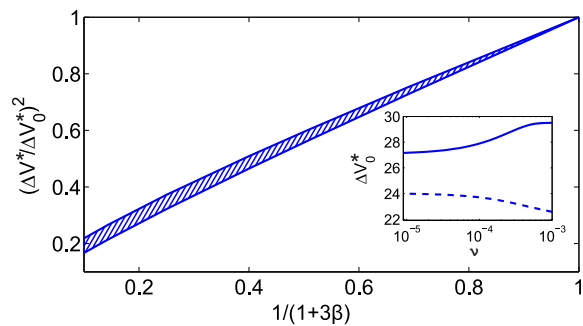}
\caption{(Color online) Squared critical potential drop, normalized to its value at $\beta=0$ vs. $1/(1+3\beta)$. The shaded region is located between $\nu=4\cdot10^{-4}$ (upper line) and $\nu=10^{-2}$ (lower line)}
\label{Crit}
\end{figure}

\begin{figure}[hbt]
\centering
\includegraphics[width=0.45\textwidth]{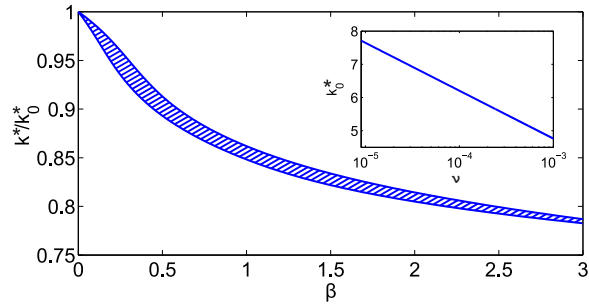}
\caption{(Color online) Critical values of the wave number normalized to its value at $\beta=0$ vs. $\beta$. The shaded region is located between $\nu=10^{-3}$ (upper line) and $\nu=10^{-2}$ (lower line)}\label{bdV}
\end{figure}

Typical marginal stability curves for different slips $\beta$ are shown in
Fig.~\ref{margi}. They clarify the effect of hydrophobicity on the
electrokinetic instability. Even a relatively small hydrophobicity $\beta=1$
results in a significant destabilization of the flow with a shift of the
threshold of instability to the left. With further increase in $\beta$, some
saturation of the influence of the hydrophobicity can be seen. The critical wavelength $2\pi/k^*$ with increasing hydrophobicity becomes larger, but this
increase is relatively weak in comparison with the changes in $\Delta V^*$. The growth
rate coefficient $\lambda(k)$ inside the instability region is presented in the
inset to the figure. A change in the hydrophobicity $\beta$ from $\beta=0$ to
$\beta=1$ is accompanied by an increase of more than ten times in $\lambda_{max}(k)$;
an obvious saturation is seen with further increase in $\beta$.

The stability results are generalized in Fig.~\ref{Crit} and Fig.~\ref{bdV}. In the
first figure, the critical potential drop $\Delta V^*$ for the hydrophobic surface
$\beta \neq 0$ is related to that for the regular no-slip surface with
$\beta=0$. The figure shows a linear behavior of $(\Delta V^*/\Delta V^*_0)^2$
with respect to $1/(1+3\beta)$, confirming the analytical prediction
of \eqref{eq61}. The numerical results, however, are located slightly above the
analytical ones and, hence, the dependence of $\Delta V^*$ on $\beta$ is weaker
than the analytics predict. The dependence on the Debye number $\nu$ is rather weak and
is characterized by the shaded region in the figure. To complete the picture,
the critical potential drop is presented as a function of the Debye number
in the inset to the figure at $\beta=0$. The solid line corresponds to the
numerics while the dashed line corresponds to the analytical formula \eqref{eq58a}. The
difference between them decreases as $\nu \to 0$, but is still finite. In
order to get a better match, the next approximation with respect to $\beta$ is
required (see \cite{Rub13}). The dependence of the critical wave number on $\beta$
is relatively weak, see Fig.~\ref{bdV}. In all our calculations, 
critical wave number decreases with increasing hydrophobicity. 

%\newpage
\textit{Nonlinear analysis.}
Recent experiments \cite{Nik1} show a strong correlation between the rate of
overlimiting current and the degree of surface hydrophobicity. In order to
capture this effect theoretically, the nonlinear behavior of the full
Nernst--Planck--Poisson--Stokes system was studied. Equations
\eqref{eq1}--\eqref{eq8a} were numerically integrated by the method described in
\cite{Dem1,Chang2} and, hence, we skip all the technical details.

\begin{figure*}[hbtp]
\centering
\includegraphics[angle=270,width=0.7\textwidth]{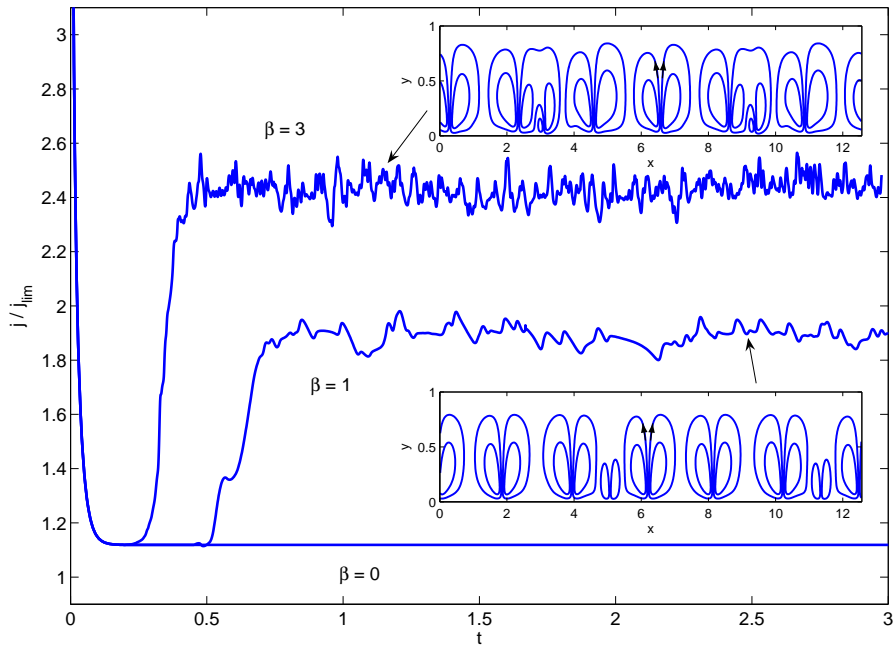}\\
\caption{(Color online) Evolution of the average electric current for $\Delta V=25$, $\varkappa=0.1$, $\nu=10^{-3}$ and $\beta=0$, $\beta=1$ and $\beta=3$. Limiting current $j_{\lim}=4$. Inset: Streamlines of electroconvective vortex pairs}
\label{Num}
\end{figure*}

A~small-amplitude white-noise spectrum is superimposed on the bulk ion
concentrations, $c^+=c^-=1$, and this superposition is taken as the intitial
conditions at $t=0$. It is convenient to present and discuss the results using
the time series $j(t)$ defined by Equation \eqref{eq9} (see Fig.~\ref{Num}). For
$\beta=0$, the 1D case is stable in small initial perturbations and the limiting
current is established after a short transition period,
$t \approx 0.1 \div 0.2$ (the order of the corresponding dimensional
transitional time is about several seconds). The surface hydrophobicity $\beta=1$
dramatically changes this behavior. Now the small-amplitude noise is increasing and
it manifests itself starting from $t \approx 0.5$. At $t \approx 0.8$, a
transition to the overlimiting currents with $j/j_{\lim} \approx 1.8$
occurs. From this time onwards, the current irregularly oscillates near its
average value with a peak-to-peak amplitude of about $0.2$. For $\beta=3$, the
overlimiting current $j/j_{\lim}$ increases up to $2.4$, the current
oscillations become more irregular, and the characteric frequency of the oscillations
increases, while their amplitude does not change much. Hence, the more
hydrophobic is the surface, the more intensive is the overlimiting current.
Qualitative comparison with the computational results of \cite{Chang2,Han} shows
that the effect of hydrophobicity is much stronger than that of surface
inhomogenity, which is in qualitative agreement with the experiments
\cite{Nik1}. 
\begin{figure}[hbtp]
\centering
\includegraphics[width=0.5\textwidth]{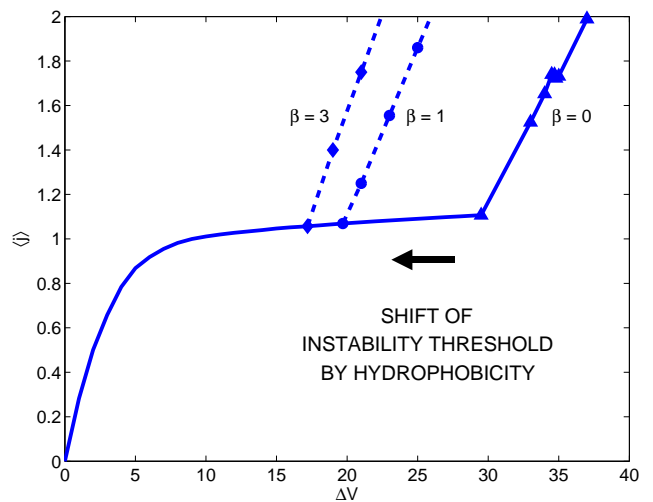}
\caption{(Color online) Volt--current characteristics at different hydrophobicities and slips $\beta$: triangles --- $\beta=0$, circles --- $\beta=1$ and diamonds ---
$\beta=3$}
\label{VC}
\end{figure}

The transition to the overlimiting regimes for $\beta=1$ and $\beta=3$ are
accompanied with a transition to a 2D flow of the liquid. Snapshots of the streamlines of
the vortex pairs are shown in the inset to the figure. While at a small
hydrophobicity, $\beta=1$, the vortex pairs behave relatively regularly, they
become more chaotic with increasing hydrophobicity, see the
snapshot for $\beta=3$.

The results of this nonlinear investigation are generalized for three values of
$\beta$ in the V--C characteristic curve presented in Fig.~\ref{VC}. Here\
$$
\langle j \rangle =\frac{1}{T}\int_{t_0}^{T+t_0}j\,dt,
$$
is the time-averaged electric current, $t_0$ is the time when the overlimiting regime
is established, and $T$ is taken large enough for averaging: $T \approx 3 \div 5$.
The overlimiting current regimes start at the points predicted by the linear
stability analysis. The manner of the dependence on the hydrophobicity is
qualitatevely similar to that predicted in experiments by Belashova et al.
\cite{Nik1} (see, for example, their figure 6(b)).

\begin{acknowledgments}
V.~S. was supported in part by the Russian Foundation for Basic Research, project No. 13-08-96536-r\_yug\_a;
E.~D. and G.~G. were supported in part by RFBR 
(projects No. 12-08-00924-a (E.~D. only), No. 14-08-31260 mol-a and No.14-08-00789-a.
E.~D. would like to thank the hospitality and support of Polymer and Crystal Physics Department
of MSU and the Head of the Department, Academician A. R. Khokhlov and Professor Olga I. Vinogradova
for fruitful comments and discussions. 
\end{acknowledgments}

%\newpage

\end{document}